\documentclass[preprint]{aastex}
\usepackage{amsmath}
\usepackage{amssymb}

\begin{document}

\title{Applications of the $k$-$\omega$ model in stellar evolutionary models}

\author{Yan Li}
\affil{Yunnan Observatories, Chinese Academy of Sciences, 
       Kunming 650216, China\\
       Key Laboratory for the Structure and Evolution of 
       Celestial Objects, Chinese Academy of Sciences\\
       University of Chinese Academy of Sciences, Beijing 100049, China\\
       Center for Astronomical Mega-Science, Chinese Academy of Sciences, 
       Beijing 100012, China}
\email{Electronic address: ly@ynao.ac.cn}

\begin{abstract}

The $k$-$\omega$ model for turbulence was first proposed by Kolmogorov (1942). A new $k$-$\omega$ model for stellar convection was developed by \citet{Li2012}, which could reasonably describe turbulent convection not only in the convectively unstable zone but also in the overshooting regions. We have revised the $k$-$\omega$ model by improving several model assumptions (including the macro-length of turbulence, convective heat flux, and turbulent mixing diffusivity, etc.), making it applicable for not only convective envelopes but also convective cores. Eight parameters are introduced in the revised $k$-$\omega$ model. It should be noted that the Reynolds stress (turbulent pressure) is neglected in the equation of hydrostatic support. We have applied it into solar models and 5$M_{\odot}$ stellar models, in order to calibrate the eight model  parameters, as well as to investigate the effects of the convective overshooting on the sun and intermediate mass stellar models. 

\end{abstract}


\section{Introduction}

The convection model is an important input physics in the studies of stellar structure and evolution. The most widely used treatment is the standard mixing-length theory \citep{bv53,bv58}. It can give the convective heat flux in the convectively unstable zone, but it cannot describe heat transfer and mixing of matter in the overshooting regions. Turbulent convection models are a better choice \citep{xio80,xio89,ca97,ca01,ly01,ly07}. They are based on the averaged moment equations of hydrodynamics, and can consider more physics of turbulence (such as generation, dissipation, diffusion, anisotropy, etc.) than the mixing-length theory (see, for example, Houdek \& Dupret 2015). Particularly, they can be used to study the properties of turbulence and the convective mixing in the overshooting regions \citep{la11a,la11b,dl14a,dl14b}.

The $k$-$\omega$ model for turbulence was first proposed by Kolmogorov (1942). Recently \citet{Li2012} developed a $k$-$\omega$ model to describe turbulent convection in stars. By solving two differential equations for the turbulent kinetic energy $k$ and the turbulence frequency $\omega$ respectively, the convective heat flux and the turbulent mixing diffusivity can be obtained not only in the convectively unstable zones but also in the overshooting regions. We applied the $k$-$\omega$ model in calculations of various stellar models, and made some improvements over the original model and the parameters it introduced. The most important improvement is that the $k$-$\omega$ model can now be applied not only in the stellar envelopes but also in the stellar cores. In the present paper, we summarize the revised $k$-$\omega$ model, and show results of some applications. In addition, eight parameters are introduced in the revised $k$-$\omega$ model. The need for calibrating these model parameters is another main aim of the present paper. It should be noted that the Reynolds stress (turbulent pressure) is neglected in the equation of hydrostatic support.

In Section 2, we give the equations of the $k$-$\omega$ model. Model assumptions are discussed and summarized in Section 3. In Section 4, we first show our results of solar models, as well as comparisons with helioseismic results, and then discuss models of a 5$M_{\odot}$ star with different treatments of the overshooting beyond the convective core and below the convective envelope. We summarize our conclusions in Section 5.

\section{ Equations of the $k$-$\omega$ model for turbulent motions }

According to \citet{Li2012}, we describe turbulent motions in stellar convection zones by two differential equations:
\begin{equation}\label{2c1}
	\frac{\partial k}{\partial t}-\frac{1}{{{r}^{2}}}\frac{\partial }{\partial r}\left( {{r}^{2}}{{\nu }_{t}}\frac{\partial k}{\partial r} \right)=G-\varepsilon 
\end{equation}
and
\begin{equation}\label{2c2}
	\frac{\partial \omega }{\partial t}-\frac{1}{{{r}^{2}}}\frac{\partial }{\partial r}\left( {{r}^{2}}\frac{{{\nu }_{t}}}{{{\sigma }_{\omega }}}\frac{\partial \omega }{\partial r} \right)=\frac{c_{L}^{2}k}{{{L}^{2}}}-{{\omega }^{2}}.
\end{equation}
In Equations (\ref{2c1}) and (\ref{2c2}), $k$ is the kinetic energy of turbulence, $\varepsilon$ the dissipation rate of $k$, and $ \omega=\varepsilon/k $ the turbulence frequency. On the left hand side of Equations (\ref{2c1}) and (\ref{2c2}), the turbulent viscosity $\nu_{t}$ is approximated by:
\begin{equation}\label{2c3}
{{\nu }_{t}}={{c}_{\mu }}\frac{{{k}^{2}}}{\varepsilon }.
\end{equation}
On the right hand side of Equation (\ref{2c1}), $G$ represents the buoyancy production rate of the turbulent kinetic energy. We omit the shear production rate in Equation (\ref{2c1}), although it can be present due to nonradial stellar pulsation \citep{dxc06,smo11}, or by meridional flows and rotation. In Equation (\ref{2c2}), $L$ represents the macro-length of turbulence, which is equivalent to the mixing-length in the standard mixing-length theory. We shall discuss these quantities in details in the subsequent sections. According to common choices \citep{pop}, the model parameters $ c_{\mu} = 0.09 $ and ${c}_{L}=c_{\mu }^{3/4}$. In Equation (\ref{2c2}), $\sigma_{\omega}$ is a model parameter.

\section{ Model assumptions }

\subsection{ Buoyancy production rate }

To determine the buoyancy production rate $G$, the full transport equations for all turbulent correlations must be used, such as given by Xiong et al. (1997) and Canuto et al. (2001). In the present paper, we adopt the so-called algebraic flux model (AFM) by imposing the local equilibrium hypothesis \citep{han93,hou99}
on all second-order moment transport equations, neglecting both the rate of change and transport terms. Then we combine the two equations for the velocity-temperature correlation $\overline{w\vartheta }$ and temperature self-correlation $\overline{{\vartheta}^2}$ to solve for the $\overline{w\vartheta }$, where $w$ is the velocity fluctuation in the vertical direction and $\vartheta$ the temperature fluctuation, and overbars stand for statistical averages for the second-order moments (for more details, see Li (2012)). According to \citet{Li2012}, we approximate the buoyancy production rate $G$ by:
\begin{equation}\label{3c11}
	G=-\frac{{{c}_{t}}}{1+{{y}^{-1}}+{{c}_{t}}{{c}_{\theta }}{{\tau }^{2}}{{N}^{2}}}\frac{{{k}^{2}}}{\varepsilon }{{N}^{2}},
\end{equation}
where $c_t$ and $c_{\theta}$ are two parameters, and $\tau=1/\omega$. It should be noted that a necessary condition for Equation (\ref{3c11}) to be valid is when 
\begin{equation}\label{3c111}
	1+{{y}^{-1}}+{{c}_{t}}{{c}_{\theta }}{{\tau }^{2}}{{N}^{2}}>0.
\end{equation}
In Equation (\ref{3c11}), the buoyancy frequency $N$ is defined by:
\begin{equation}\label{3c12}
{{N}^{2}}=\beta g\frac{T}{{{c}_{p}}}\frac{\partial s}{\partial r},
\end{equation}
and
\begin{equation}\label{3c13}
y=\frac{\rho {{c}_{p}}}{\lambda }\frac{{{k}^{2}}}{\varepsilon }.
\end{equation}
In Equations (\ref{3c12}) and (\ref{3c13}), $\rho$ is the density, $T$ the temperature, $p$ the pressure, $s$ the entropy, $c_p$ the specific heat at constant pressure, and $g$ the gravitational acceleration. In addition, the thermodynamic coefficient $\beta$ is defined by
\begin{equation}\label{3c14}
	\beta =-\frac{1}{\rho }{{\left( \frac{\partial \rho }{\partial T} \right)}_{p}},
\end{equation}
and the radiation diffusivity $\lambda$ is defined by
\begin{equation}\label{3c15}
\lambda =\frac{16\sigma {{T}^{3}}}{3\rho \kappa },
\end{equation}
where $\kappa$ is the Rosseland mean opacity and $\sigma$ is the Stefan-Boltzmann constant.

In a chemically homogeneous region, the buoyancy frequency $N$ can be expressed as:
\begin{equation}\label{3c16}
{{N}^{2}}=-\frac{\beta gT}{{{H}_{p}}}\left( \nabla -{{\nabla }_{ad}} \right),
\end{equation}
where the thermodynamic coefficient $\nabla_{ad}$ is known as the adiabatic temperature gradient, and the local temperature gradient $\nabla$ is defined as:
\begin{equation}\label{3c17}
\nabla =\frac{d\ln T}{d\ln p}.
\end{equation}
In Equation (\ref{3c16}), the local pressure scale height $H_p$ is defined in the hydrostatic equilibrium state as:
\begin{equation}\label{3c18}
{{H}_{p}}=-\frac{dr}{d\ln p}.
\end{equation}

\subsection{ Convective heat flux }

The convective heat flux is defined by:
\begin{equation}\label{3c20}
{{F}_{C}}=\rho {{c}_{p}} \overline{w\vartheta }.
\end{equation}
As pointed out in \citet{Li2012} and \citet{hou99}, the convective heat flux $F_C$ is essentially related to the buoyancy production rate $G$. We approximate the convective heat flux $F_C$ as:
\begin{equation}\label{3c21}
	{{F}_{C}}=-\frac{\rho {{c}_{p}}}{\beta g}\frac{{{c}_{h}}{{c}_{q}}}{1+{{y}^{-1}}+{{c}_{t}}{{c}_{\theta }}{{\tau }^{2}}{{N}^{2}}}\frac{{{k}^{2}}}{\varepsilon }{{N}^{2}},
\end{equation}
where $c_h$ is a parameter, and $c_q$ is based on Kays-Crawford model \citep{KC1993,WFC1997}:
\begin{equation}\label{3c22}
	{{c}_{q}}=\frac{1}{2}+c_{L}^{2}y-{{\left( c_{L}^{2}y \right)}^{2}}\left( 1-{{e}^{-1/c_{L}^{2}y}} \right).
\end{equation}
An important feature of the present convective heat flux model is that it ensures positive convective heat flux in the convectively unstable zone ($\nabla > \nabla_{ad}$) while gives negative convective heat flux in the overshooting regions ($\nabla < \nabla_{ad}$).

In a stellar convection zone, heat will be usually transferred outwardly by radiation and convection. If we denote the total heat flux by $F$, then we have that
\begin{equation}\label{3c23}
F=\frac{\lambda T}{{{H}_{p}}}{{\nabla }_{r}}={{F}_{C}}-\lambda \frac{\partial T}{\partial r},
\end{equation}
where $\nabla_r$ is known as the radiative temperature gradient. It should be noticed that the simple diffusion approximation for the radiative heat flux is only valid in the deep stellar interior, and breaks down in the stellar atmosphere. Substituting Equation (\ref{3c21}) into Equation (\ref{3c23}) and using Equation (\ref{3c16}), the temperature gradient in the stellar convection zone can be expressed as:
\begin{equation}\label{3c24}
h=\frac{{{\nabla }_{r}}-\nabla }{\nabla -{{\nabla }_{ad}}}=\frac{{{c}_{h}}{{c}_{q}}y}{1+{{y}^{-1}}+{{c}_{t}}{{c}_{\theta }}{{\tau }^{2}}{{N}^{2}}}.
\end{equation}
Besides, Equation (\ref{3c16}) can be rewritten as:
\begin{equation}\label{3c25}
{{N}^{2}}=-\frac{E}{1+h},
\end{equation}
where 
\begin{equation}\label{3c26}
E=\frac{\beta gT}{{{H}_{p}}}\left( {{\nabla }_{r}}-{{\nabla }_{ad}} \right).
\end{equation}

\subsection{ Macro-length of turbulence }

In the standard mixing-length theory, the macro-length of turbulence is usually assumed to be proportional to the local pressure scale height (as discussed in \citet{Li2012}):
\begin{equation}\label{3c31}
L={{c}_{L}}\alpha {{H}_{p}},
\end{equation}
where $\alpha$ is the so-called mixing-length parameter. This model is mostly appropriate for turbulence in the convective envelopes of stars, where the thickness of the stellar convective envelope is usually much larger than the local pressure scale height. 

In the stellar convective cores, however, the radii of the convective cores are usually smaller than the local pressure scale height, which significantly restricts the development of turbulence. Accordingly, the macro-length of turbulence will therefore be restricted practically by the thickness of the convection zones. We therefore suggest a model for the macro-length of turbulence in the convective core that
\begin{equation}\label{3c32}
L={{c}_{L}}{\alpha }'D,
\end{equation}
where $D$ is the radius of the convective core, and ${\alpha }'$ is a parameter similar as introduced before.

\subsection{ Formulation of the $k$-$\omega$ model for stellar convection}

Using Equations (\ref{3c24}) and (\ref{3c25}) we obtain that
\begin{equation}\label{5c11}
\left( 1+{{y}^{-1}} \right){{h}^{2}}+\left( 1+{{y}^{-1}}-{{c}_{t}}{{c}_{\theta }}{{\tau }^{2}}E-{{c}_{h}}{{c}_{q}}y \right)h-{{c}_{h}}{{c}_{q}}y=0.
\end{equation}
The solution of Equation (\ref{5c11}) can be written as:
\begin{equation}\label{5c12}
h=\frac{\sqrt{{{\left( 1+{{y}^{-1}}-{{c}_{t}}{{c}_{\theta }}{{\tau }^{2}}E-{{c}_{h}}{{c}_{q}}y \right)}^{2}}+4{{c}_{h}}{{c}_{q}}y\left( 1+{{y}^{-1}} \right)}-\left( 1+{{y}^{-1}}-{{c}_{t}}{{c}_{\theta }}{{\tau }^{2}}E-{{c}_{h}}{{c}_{q}}y \right)}{2\left( 1+{{y}^{-1}} \right)}.
\end{equation}
Substituting Equations (\ref{5c12}) and (\ref{3c25}) into Equation (\ref{3c11}), we finally obtain that
\begin{equation}\label{5c13}
G=\frac{2{{c}_{t}}E}{A+\sqrt{{{A}^{2}}+4{{c}_{h}}{{c}_{q}}y\left( {{c}_{t}}{{c}_{\theta }}{{\tau }^{2}}E \right)}}\frac{k}{\omega },
\end{equation}
where
\begin{equation}\label{5c14}
A=\left( 1+{{y}^{-1}}+{{c}_{h}}{{c}_{q}}y-{{c}_{t}}{{c}_{\theta }}{{\tau }^{2}}E \right).
\end{equation}

Using Equations (\ref{3c31}), (\ref{3c32}), and (\ref{5c13}), we rewrite the equations of the $k$-$\omega$ model for stellar convection as:
\begin{equation}\label{5c15}
	\frac{\partial k}{\partial t}-\frac{\partial }{\partial m}\left( {{\left( 4\pi \rho {{r}^{2}} \right)}^{2}}{{\nu }_{t}}\frac{\partial k}{\partial m} \right)=\frac{2{{c}_{t}}E}{A+\sqrt{{{A}^{2}}+4{{c}_{h}}{{c}_{q}}y\left( {{c}_{t}}{{c}_{\theta }}{{\tau }^{2}}E \right)}}\frac{k}{\omega }-k\omega
\end{equation}
and
\begin{equation}\label{5c16}
	\frac{\partial \omega }{\partial t}-\frac{\partial }{\partial m}\left( {{\left( 4\pi \rho {{r}^{2}} \right)}^{2}}\frac{{{\nu }_{t}}}{{{\sigma }_{\omega }}}\frac{\partial \omega }{\partial m} \right)=\frac{c_{L}^{2}k}{{{L}^{2}}}-{{\omega }^{2}},
\end{equation}
where $m$ is the mass within the sphere of radius $r$. 

Setting up boundary conditions for Equations (\ref{5c15}) and (\ref{5c16}) are usually a simple matter. In most cases, overshooting regions are supposed to exist outside a convection zone. As pointed out by Xiong (1985), Freytag et al. (1996), and Asida \& Arnett (2000), the turbulent velocity field decays exponentially beyond the convective boundary. This property makes the choice of the boundary conditions an easy task. However, the boundary conditions should be properly set up in some special cases (for example, at the stellar center), in order to get the correct solutions of Equations (\ref{5c15}) and (\ref{5c16}). 

In practice, we use a numerical scheme similar as presented in \citet{Li2012} to solve the equations of the $k$-$\omega$ model. An independent Fortran module is developed to solve the equations of the $k$-$\omega$ model by use of the Newton iterative method. In the convective regions, we can obtain $y$ by solving Equations (\ref{5c15}) and (\ref{5c16}). Afterwards, using Equation (\ref{3c24}) we can obtain the temperature gradient by:
\begin{equation}\label{5c17}
\nabla =\frac{h}{1+h}{{\nabla }_{ad}}+\frac{1}{1+h}{{\nabla }_{r}}.
\end{equation}

\subsection{ Turbulent mixing flux of elements }

Stellar convection can transport not only heat but also matter. When an amount of matter moves from a higher temperature layer to a lower temperature layer, the same amount of matter must move simultaneously from the lower temperature layer to the higher temperature layer in order to keep the conservation of matter . It is evident that the exchange of matter is accompanied by the transport of heat.

Usually in turbulence models for stellar convection, the density fluctuation can be decomposed (for example, \citet{zh2013}) as:
\begin{equation}\label{6c11}
	{\rho }'={{\left( \frac{\partial \rho }{\partial T} \right)}_{p,{{X}_{i}}}}{T}'+\sum\limits_{i}{{{\left( \frac{\partial \rho }{\partial {{X}_{i}}} \right)}_{p,T}}{{{{X}'}}_{i}}},
\end{equation}
where $X_i$ is the mass fraction of element "$i$" and a prime denotes the corresponding turbulent fluctuation of a physical quantity. In Equation (\ref{6c11}), the summation should be carried out for all considered species. Accordingly, the transport effects of turbulence can be described by:
\begin{equation}\label{6c12}
	\overline{w{\rho }'}={{\left( \frac{\partial \rho }{\partial T} \right)}_{p,{{X}_{i}}}}\overline{w{T}'}+\sum\limits_{i}{{{\left( \frac{\partial \rho }{\partial {{X}_{i}}} \right)}_{p,T}}\overline{w{{{{X}'}}_{i}}}},
\end{equation}
where $w$ is the turbulent velocity in the vertical direction. It can be noticed in Equation (\ref{6c12}) that the term containing $\overline{w{T}'}$ is related to the turbulent heat flux, while the term containing $\overline{w{{{{X}'}}_{i}}}$ is related to the flux of turbulent mixing. As the two processes come from the same movements of turbulence, their efficiencies should be related with each other.

Based on above arguments, we assume that the efficiency of turbulent mixing is proportional to the efficiency of turbulent heat transport. Therefore, referring to the form of Equation (\ref{3c21}),  we approximate the turbulent mixing flux $F_{X,i}$ of element $i$ as:
\begin{equation}\label{6c13}
	{{F}_{X,i}}=\overline{w{{{{X}'}}_{i}}}=-\frac{{{c}_{X}}}{1+{{y}^{-1}}+{{c}_{t}}{{c}_{\theta }}{{\tau }^{2}}{{N}^{2}}}\frac{{{k}^{2}}}{\varepsilon }\frac{\partial {{X}_{i}}}{\partial r},
\end{equation}
where $c_X$ is a parameter. Our model of Equation (\ref{6c13}) is similar to what was proposed by \citet{zh2013} for the convective overshooting mixing.

Using Equation (\ref{6c13}), the equation of element evolution in the spherical symmetry can be written as:
\begin{equation}\label{6c14}
	\frac{\partial {{X}_{i}}}{\partial t}+\frac{\partial }{\partial m}\left( 4\pi \rho {{r}^{2}}{w_{i}}{{X}_{i}} \right)=\frac{\partial }{\partial m}\left( {{\left( 4\pi \rho {{r}^{2}} \right)}^{2}}{{D}_{t}}\frac{\partial {{X}_{i}}}{\partial m} \right)+{d_i},
\end{equation}
where $w_{i}$ is the diffusion velocity of element "$i$" given by, for example, Thoul et al. (1994), and $d_i$ the generation rate of element "$i$" According to Equation (\ref{6c13}), we define the turbulent diffusivity $D_t$ as:
\begin{equation}\label{6c15}
	{{D}_{t}}=\frac{{{c}_{X}}}{1+{{y}^{-1}}+{{c}_{t}}{{c}_{\theta }}{{\tau }^{2}}{{N}^{2}}}\frac{k}{\omega }.
\end{equation}
Our treatment is basically similar with that of Langer et al. (1985), except that we use Equation (\ref{6c15}) to calculate the diffusivity $D_t$.

\begin{deluxetable}{cccccccc}
\tablewidth{0pt}
\tablecaption{ Parameters of the $k$-$\omega$ model}
\tablehead{ \colhead{$c_{\mu}$} &    \colhead{$\sigma_{\omega} $} & 
            \colhead{$c_{\theta}$} & \colhead{$c_t$} & 
            \colhead{$c_h$} &        \colhead{$c_X$} &          
            \colhead{$\alpha$} & \colhead{${\alpha}'$}  }

\startdata
0.09 & 1.5 & 0.5 & 0.1 & 2.344 & 1.0 & 0.7 & 0.06 \\
\enddata
\end{deluxetable}

\begin{deluxetable}{lcccccc}
\tablecaption{ Parameters of solar models }
\tablewidth{0pt}
\tablehead{ \colhead{Model}      &    \colhead{$c_h$} & 
            \colhead{$Y_0$}      &    \colhead{$Y_s$} & 
            \colhead{$Z_0$}      & \colhead{$Z/X$}    & 
            \colhead{ $R_{CZ}/R_{\odot}$ }}
\startdata
MLT & 2.413 & 0.27312 & 0.24312 & 0.02 & 0.024371 & 0.7137 \\
OSM & 2.351 & 0.27315 & 0.25126 & 0.02 & 0.025622 & 0.7143 \\
\enddata
\end{deluxetable}

\section{ Applications of the $k$-$\omega$ model }

The $k$-$\omega$ model proposed in the previous sections contains eight parameters, i.e., $c_{\mu}$, $ \sigma_{\omega} $, $c_{\theta}$, $c_t$, $c_h$, $c_X$, $\alpha$, and ${\alpha}'$.  Values of some parameters have already been given in the literature \citep{pop}. However, we have to determine values of the other parameters. Therefore, we try to incorporate the $k$-$\omega$ model into calculations of stellar evolution models, in order to investigate the effects of the parameters on the turbulent convection model itself, as well as on the stellar structure and evolution. We first introduce the input physics and numerical schemes of our calculations, and then show our results and give some discussions.  

\subsection{ Input physics and numerical schemes }

Our stellar models were computed by a stellar evolution code h04.f, which was originally described by Paczynski and Kozlowski and updated by \citet{sie2004}. The nuclear reaction rates were updated according to \citet{bap} and \citet{har}. We updated opacity tables and the equation of state data according to Zhang (2015). In particular, EOS2005 tables \citep{RN2002} were used, and radiative opacities from \citet{Ferguson2005} and OPAL \citep{IR1996} were adopted. We modified the treatment of element diffusion according to \citet{tbl}. The standard mixing-length theory is used in the convection zones, except for particularly specified. 

Values of parameters adopted in the $k$-$\omega$ model are listed in Table 1, unless specified otherwise. We implemented the $k$-$\omega$ model into the stellar evolution code as follows. When a stellar model was obtained by solving the equations of the stellar structure, we solved then Equations (\ref{5c15}) and (\ref{5c16}) to obtain $\nabla$ and $D_t$ by use of Equations (\ref{5c17}) and (\ref{6c15}). We applied $D_t$ in Equation (\ref{6c14}) to obtain the chemical profile in the stellar interior for the following time step, and then used $\nabla$ given by Equation (\ref{5c17}) to solve the equations of the stellar structure. If a better accuracy is required, the equations of both the stellar structure and the $k$-$\omega$ model can be iteratively solved to obtain a self-consistent solution. 

\subsection{ Solar models }

Our solar models were computed by use of the $k$-$\omega$ model (e.g. Equations (\ref{5c15}) and (\ref{5c16})), which will be referred to in this paper as the overshooting solar model (OSM). For comparisons, we also computed the standard solar model (SSM) by use of the mixing-length theory. The stellar models were evolved from the zero age main-sequence to the present solar age of 4.57\,Gyr, to achieve $R_{\odot}=6.9566\times{10^{10}}$cm  \citep{hab} and $L_{\odot}=3.839\times{10^{33}}$erg/s \citep{bap} with an accuracy of $10^{-4}$. The element diffusion was included in our solar models.

Our standard solar models were computed according the mixing-length theory to treat convection in the stellar envelope. Instead of using the original form of the standard mixing-length theory, we used the local equilibrium solutions of Equations (1) and (2), which reduce to a cubic equation exactly the same form as the standard mixing-length theory but with different definitions of model parameters. As a result, the adjustable parameter is $c_h$ instead of the original mixing-length parameter $\alpha$. We adjusted the initial helium abundance $Y_0$ and the value of parameter $c_h$ iteratively to achieve those solar values. In order to obtain a correct radius at the base of the convective envelope according to the result of helioseismic inversion $R_{CZ}=0.713R_{\odot}$ \citep{ch1991, BA1997}, we had to adopt an initial metal abundance $Z_0=0.02$, which resulted in a higher $Z/X$ (0.0244, the standard solar model) compared to the recent observations (0.0181, \citet{asp}). 

For the overshooting solar model, we used the solution of the $k$-$\omega$ model to compute not only the overshooting mixing but also the temperature gradient in the solar convective envelope. Similarly we adjusted  $Y_0$ and $c_h$ iteratively to calibrate the stellar model achieving the solar values. The basic properties of our solar models are summarized in Table 2.

\begin{figure}
\plotone{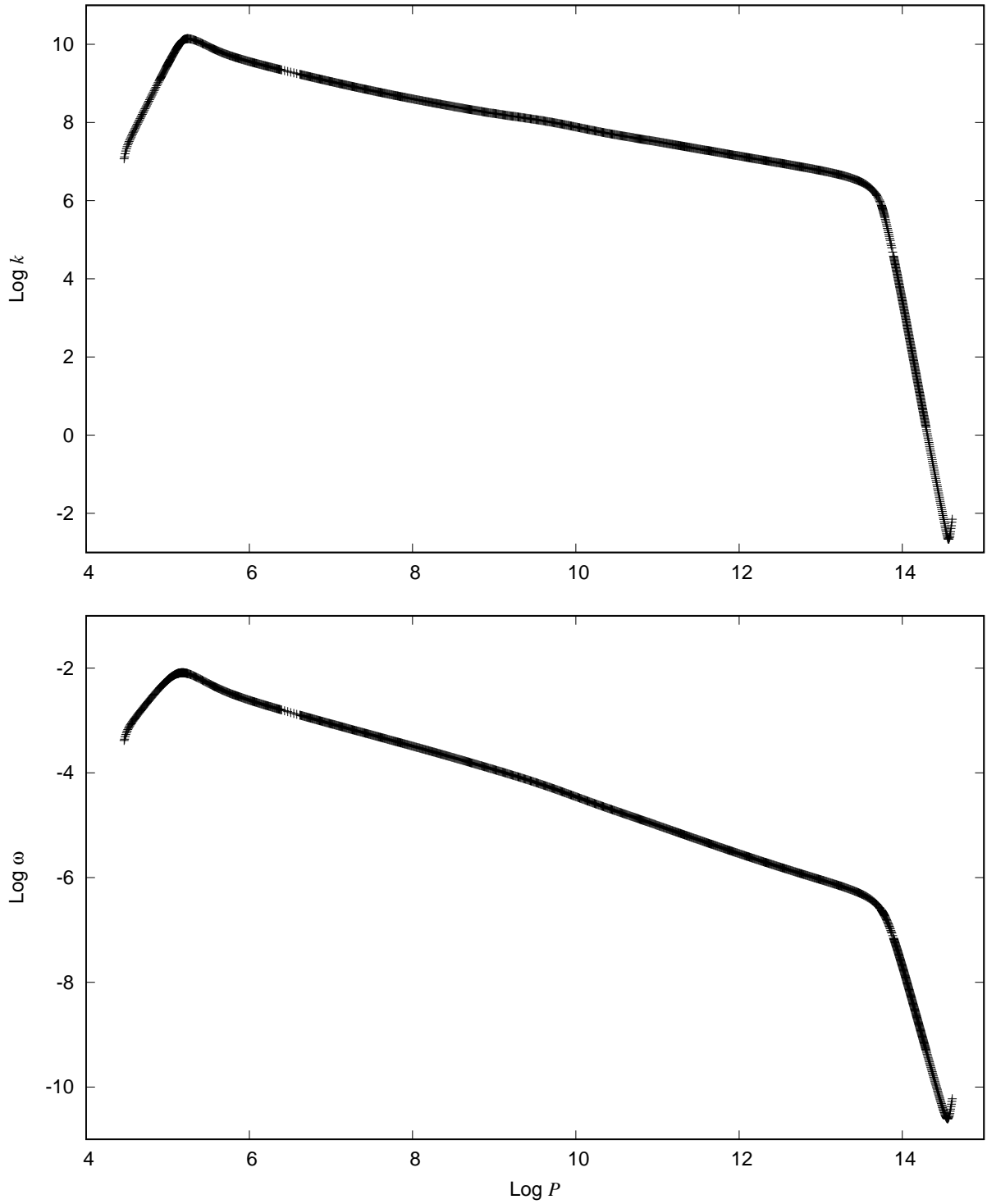}
\caption{Distributions of the turbulent kinetic energy (upper panel) and turbulence frequency (lower panel) for the overshooting solar model. }
\end{figure}

\begin{figure}
\plotone{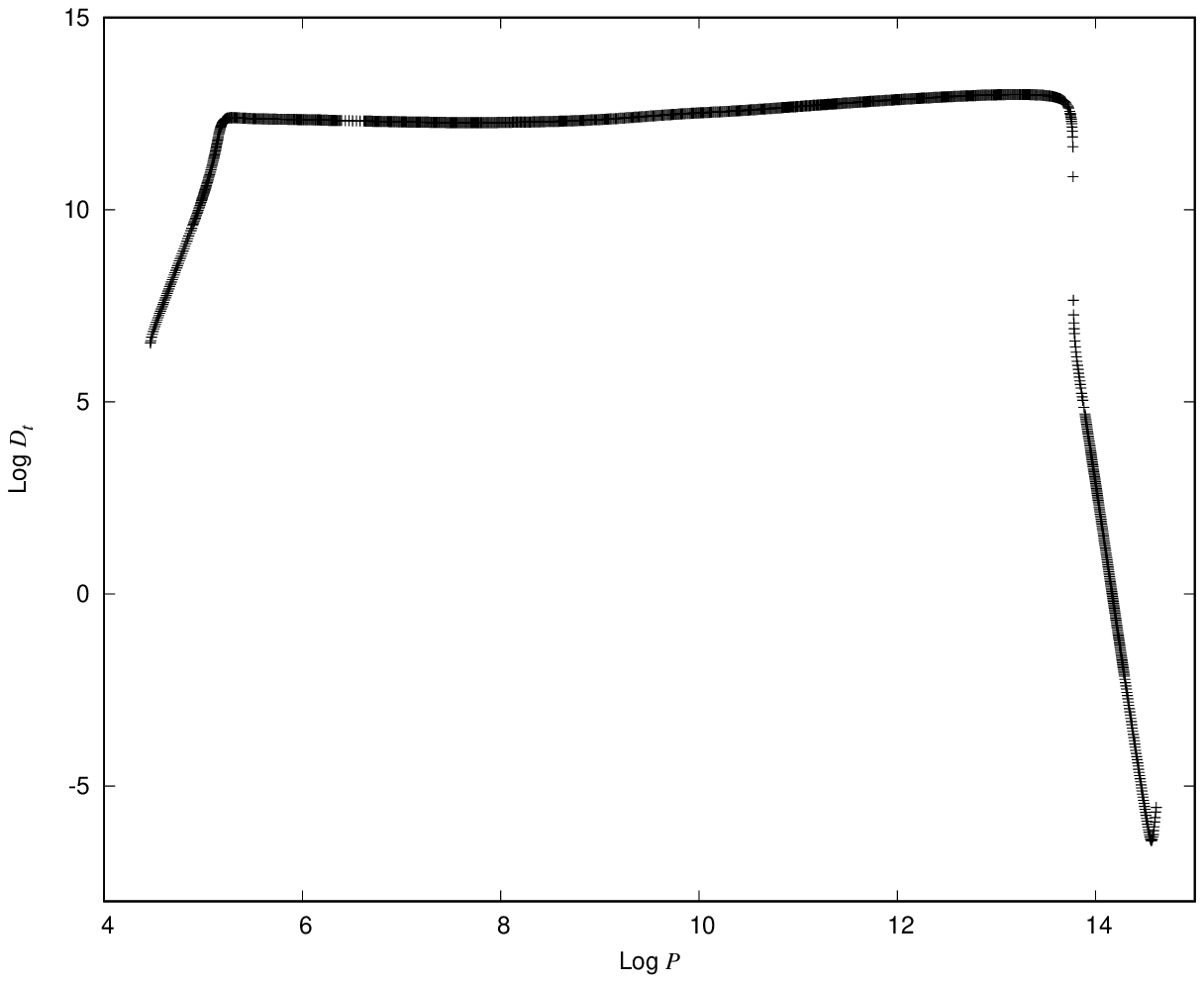}
\caption{Distribution of the turbulent diffusivity for the overshooting solar model. }
\end{figure}

The distributions of turbulent kinetic energy and turbulence frequency are shown in Figure 1 for the overshooting solar model. It can be seen that there are two overshooting regions in the solar convective envelope. The lower one is located just below the base of the convection zone, which is responsible for the overshooting mixing below the base of the convection zone. However, the upper one is developed below the top of the convection zone and has been extending outward into the solar atmosphere. The turbulent diffusivity is given in Figure 2 for the solar convective envelope. We can find that turbulence decays rapidly below the base of the convection zone, resulting in an effective mixing distance of about $1H_p$. On the other hand, the overshooting develops so efficiently into the atmosphere that it can overwhelm the gravitational settling effect to insure a complete mixing in the solar atmosphere. 

\begin{figure}
\plotone{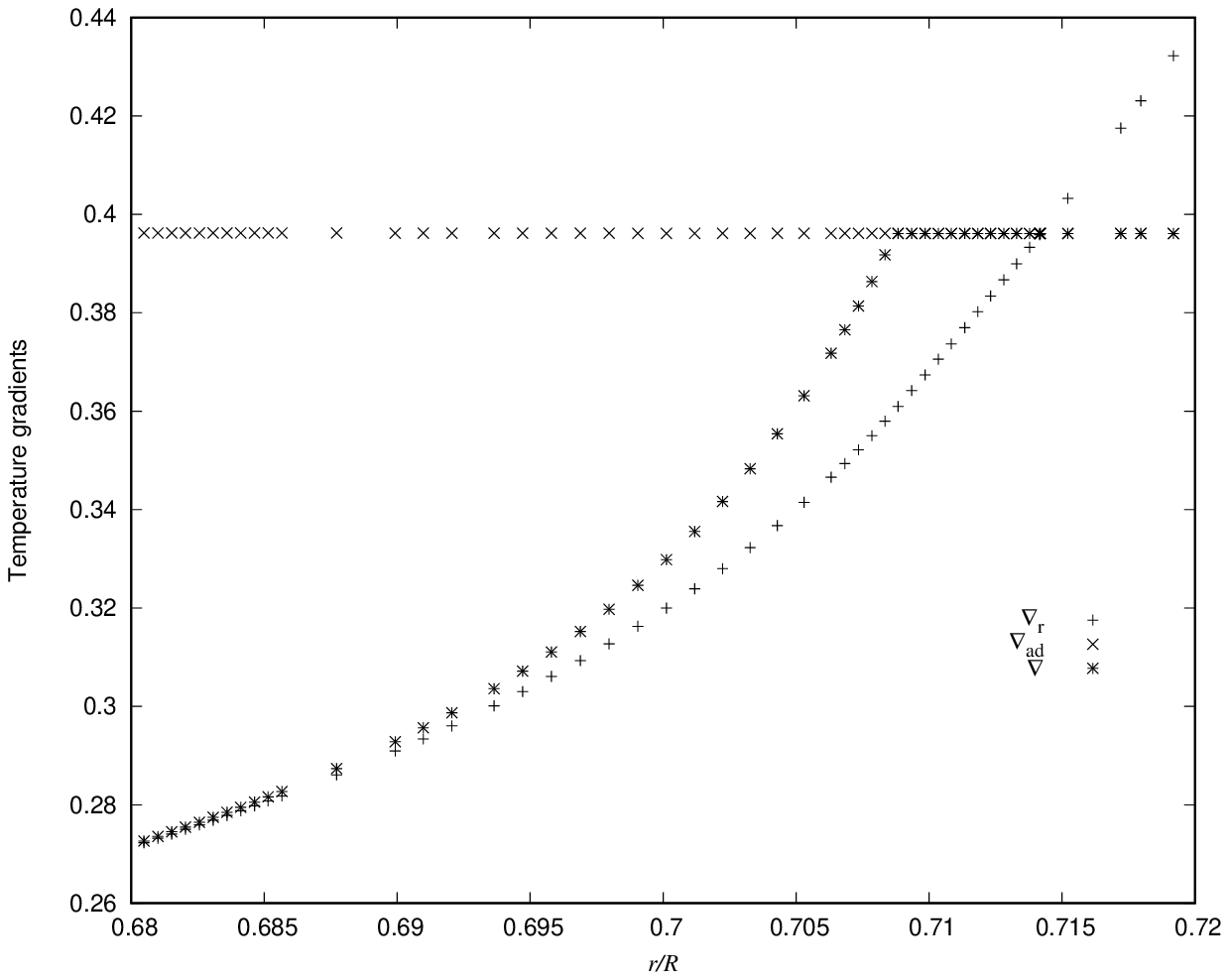}
\caption{Distributions of the temperature gradient, the radiative temperature gradient, and the adiabatic temperature gradient near the base of the convective envelope for the overshooting solar model. }
\end{figure}

\begin{figure}
\plotone{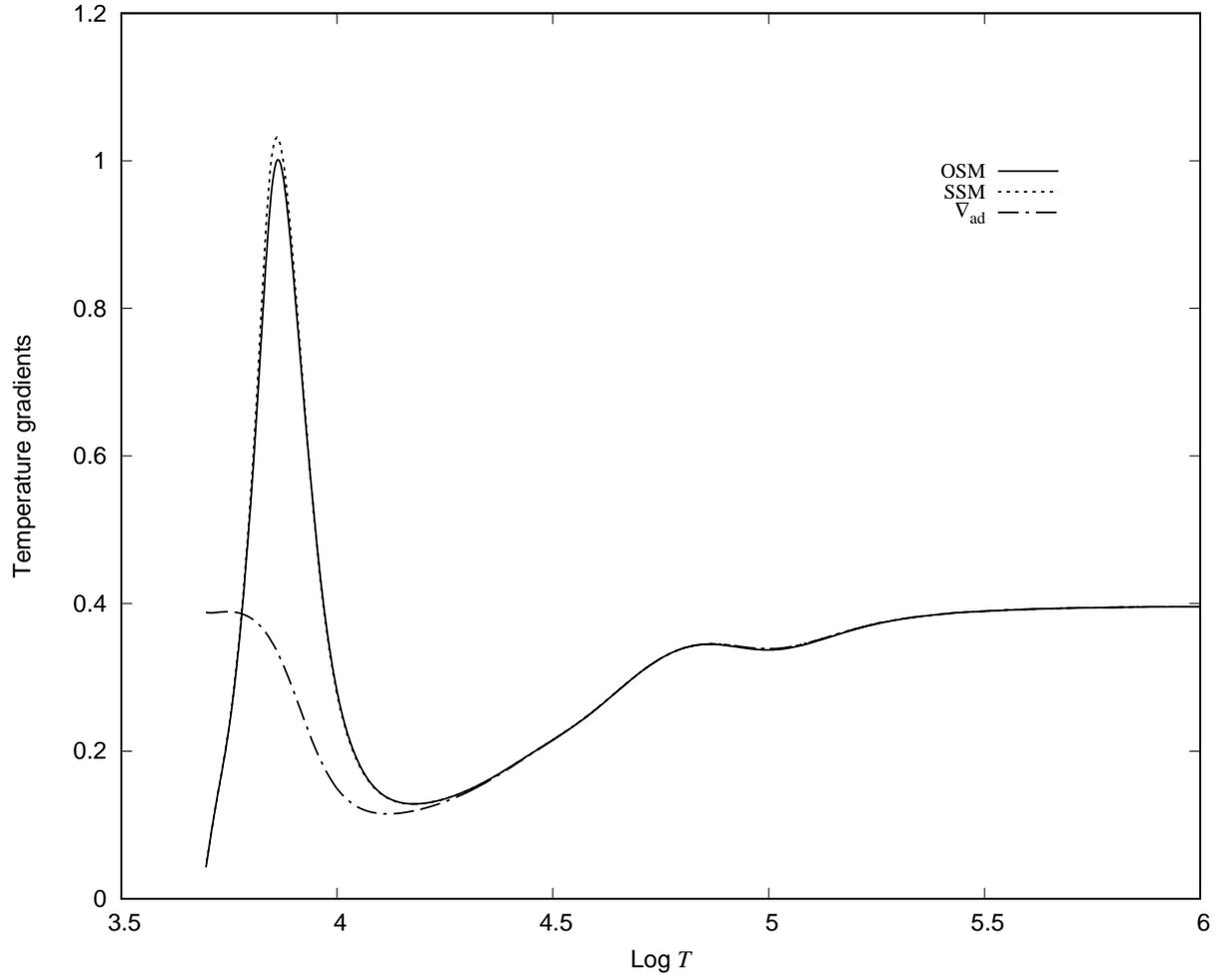}
\caption{Distributions of the temperature gradient and the adiabatic temperature gradient for the standard and overshooting solar models. }
\end{figure}

\begin{figure}
\plotone{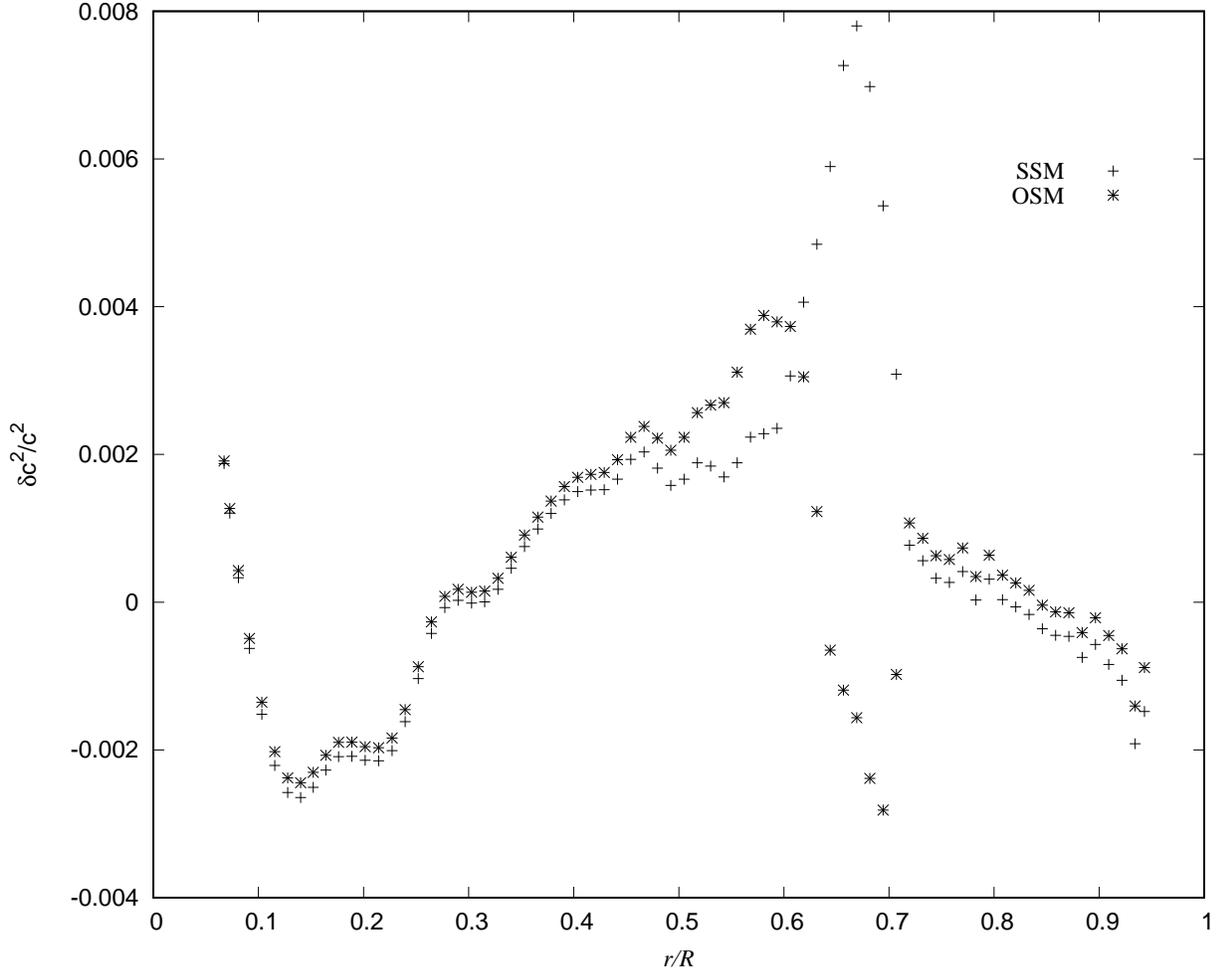}
\caption{Differences of the sound speeds between helioseismic result and prediction of the solar model for the standard and overshooting solar models. }
\end{figure}

\begin{figure}
\plotone{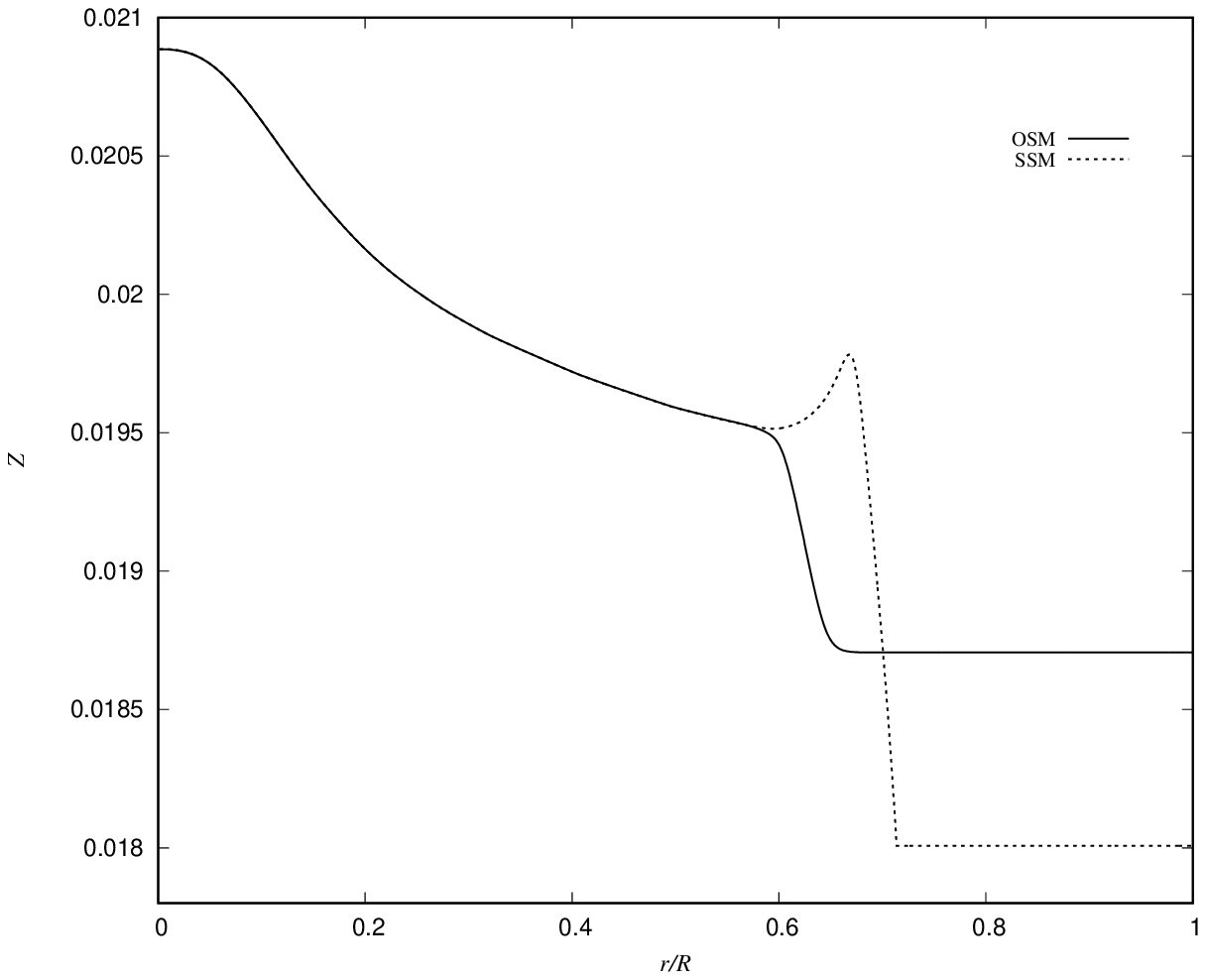}
\caption{Distributions of the metal abundance for the standard and overshooting solar models. }
\end{figure}

The distributions of different temperature gradients below the base of the convection zone are shown in Figure 3 for the overshooting solar model. It can be seen that there is a sub-adiabatic but over-radiative temperature gradient zone just below the base of the convection zone, which indicates that the convective heat flux is negative in this region. It consists of a shorter nearly adiabatic part (about 0.005 $R_{\odot}$) and a longer sub-adiabatic part (about 0.025 $R_{\odot}$). It can be found from our result that a nearly adiabatic overshooting zone corresponds roughly to a completely mixing region, and these results support the commonly used mixing-length type overshooting approach. However, the linearly logarithmic decay of the turbulent diffusivity in Figure 2 is basically similar to the overshooting-mixing model of \citet{Fr2012}. In addition, the presence of a sub-adiabatic but over-radiative temperature gradient zone below the base of the solar convection zone also satisfies the restriction of helioseismological investigations \citep{CD2011}. On the other hand, the temperature gradients are compared around the photosphere for the standard and overshooting solar models in Figure 4. It can be seen that they show similar distributions, although the peak of the temperature gradient for the overshooting solar model is a little bit lower than that of the standard solar model. 

The sound speed differences between solar models and the helioseismic inversion \citep{BPB1999} are shown in Figure 5, and distributions of metal abundance for the standard and overshooting models are shown in Figure 6. It can be seen that our results agree well with \citet{zh2015}, i.e., the metal abundance bump just below the base of the convection zone in the standard solar model is responsible for the peak of the sound speed difference in Figure 5. Such a metal abundance bump will result in a higher mean molecular weight and hence a lower sound speed there. It is worth to note that the overshooting mixing effectively operates in a distance of about 0.13$R_{\odot}$, which is about 5 times longer than the overshooting effect on the temperature gradient as already shown in Figure 3 \citep{ZL2012}.

What are the effects of parameters of the $k$-$\omega$ model on the convective overshooting in the solar model? The most important parameter is the mixing-length parameter $\alpha$, which determines not only the length of adiabatic overshooting region just below the base of the convection zone but also the slope that the turbulent kinetic energy decays in both overshooting regions. Numerical experiments show that the bigger the value of $\alpha$ is, the smaller the spatially decaying slope of the turbulent kinetic energy and the longer the adiabatic overshooting zone will be in the overshooting region. The mixing-length parameter $\alpha$ also has a significant effect on the turbulent heat flux, and the bigger the value of $\alpha$, the higher the turbulent heat flux. This effect is most significant just below the photosphere in the overshooting solar model, and the smaller the value of $\alpha$, the higher the peak of the temperature gradient in Figure 4. The next important parameter is $c_h$, which directly determines the turbulent heat flux. As a result, a smaller value of $\alpha$ can be compensated by a larger value of $c_h$ to keep the turbulent heat flux almost unchanged. In addition, $c_X$ controls the efficiency of the overshooting mixing. 

\subsection{ Overshooting in a $5M_{\odot}$ star }     

Evolutionary models were computed for a $5M_{\odot}$ star. We adopted the initial helium abundance $Y=0.276$ and initial metal abundance $Z=0.020$, and set parameters $c_h=2.344$ and $c_X=1.0$. The star developed a convective core during the main sequence, and we included the overshooting mixing process beyond the convective core. Numerical experiments showed that the distance of the overshooting mixing sensitively depended on the value of the parameter ${\alpha}'$, and we adopted ${\alpha}'=0.06$. When the star evolved onto the red giant branch, it developed a convective envelope. Then we included the overshooting mixing below the base of the convective envelope, and adopted ${\alpha}=0.70$ as in the solar case. However, we ignored the effect of convective overshooting on the temperature gradient, for it had been shown to be quite small in the overshooting solar model. In addition, we did not consider the element diffusion here.

\begin{figure}
\plotone{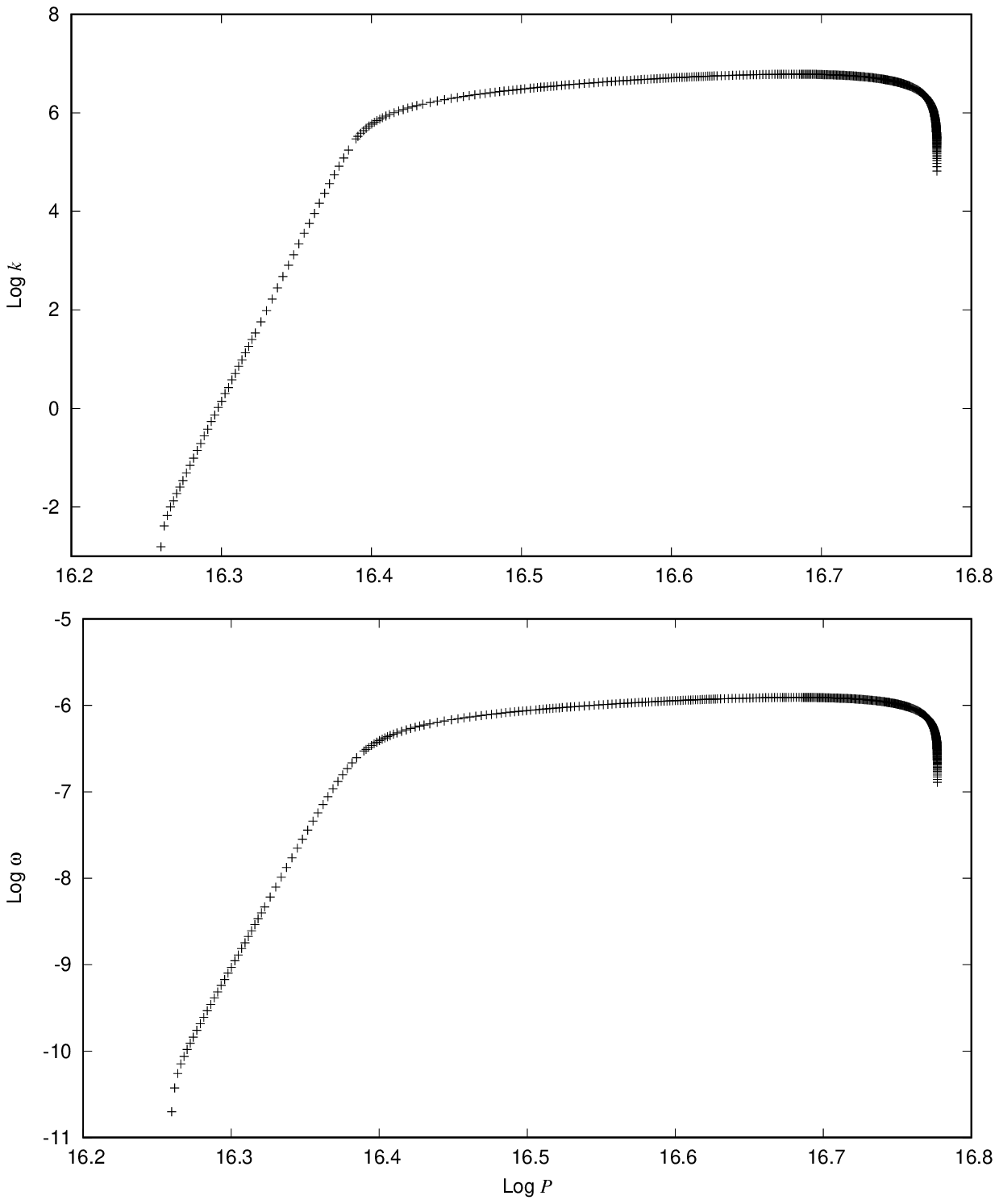}
\caption{Distributions of the turbulent kinetic energy (upper panel) and turbulence frequency (lower panel) for a 5$M_{\odot}$ main sequence model. }
\end{figure}

\begin{figure}
\plotone{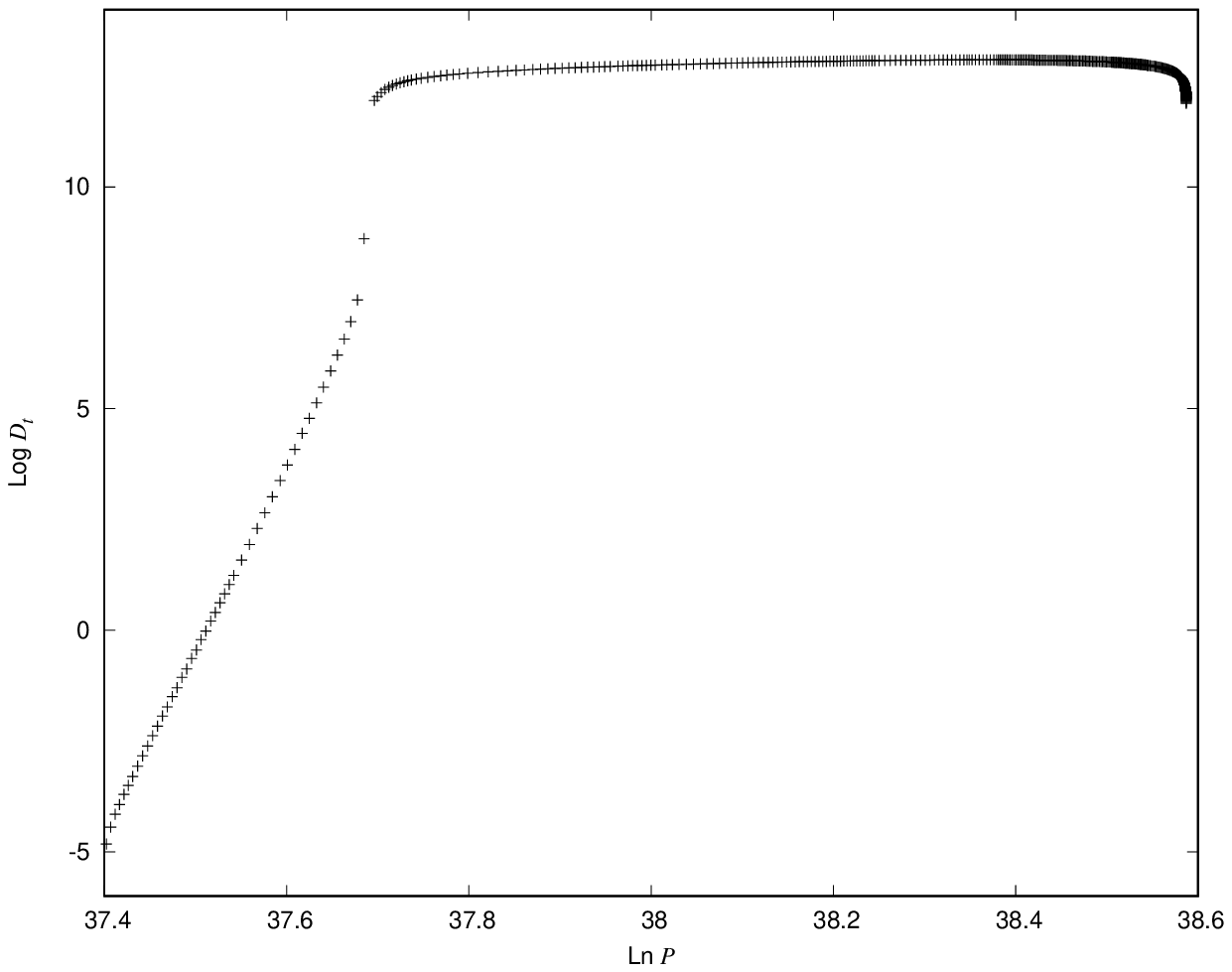}
\caption{Distribution of the turbulent diffusivity for a 5$M_{\odot}$ main sequence model. }
\end{figure}

Typical distributions of the turbulent kinetic energy and the turbulence frequency are shown in Figure 7 for an overshooting stellar model (OSM) during the main sequence. It can be seen that turbulence still decays linearly in the overshooting region, but with a slope much larger than in the cases of the solar overshooting regions. The distribution of the turbulent diffusivity is given in Figure 8. It can be noticed that the convective overshooting results in an effective mixing distance of about $0.2H_p$ beyond the convective core. The distributions of hydrogen abundance are compared for the standard and overshooting stellar models in Figure 9. It can be seen that the convective core overshooting enlarges the mass of the completely mixing core by about 20 percent.

\begin{figure}
\plotone{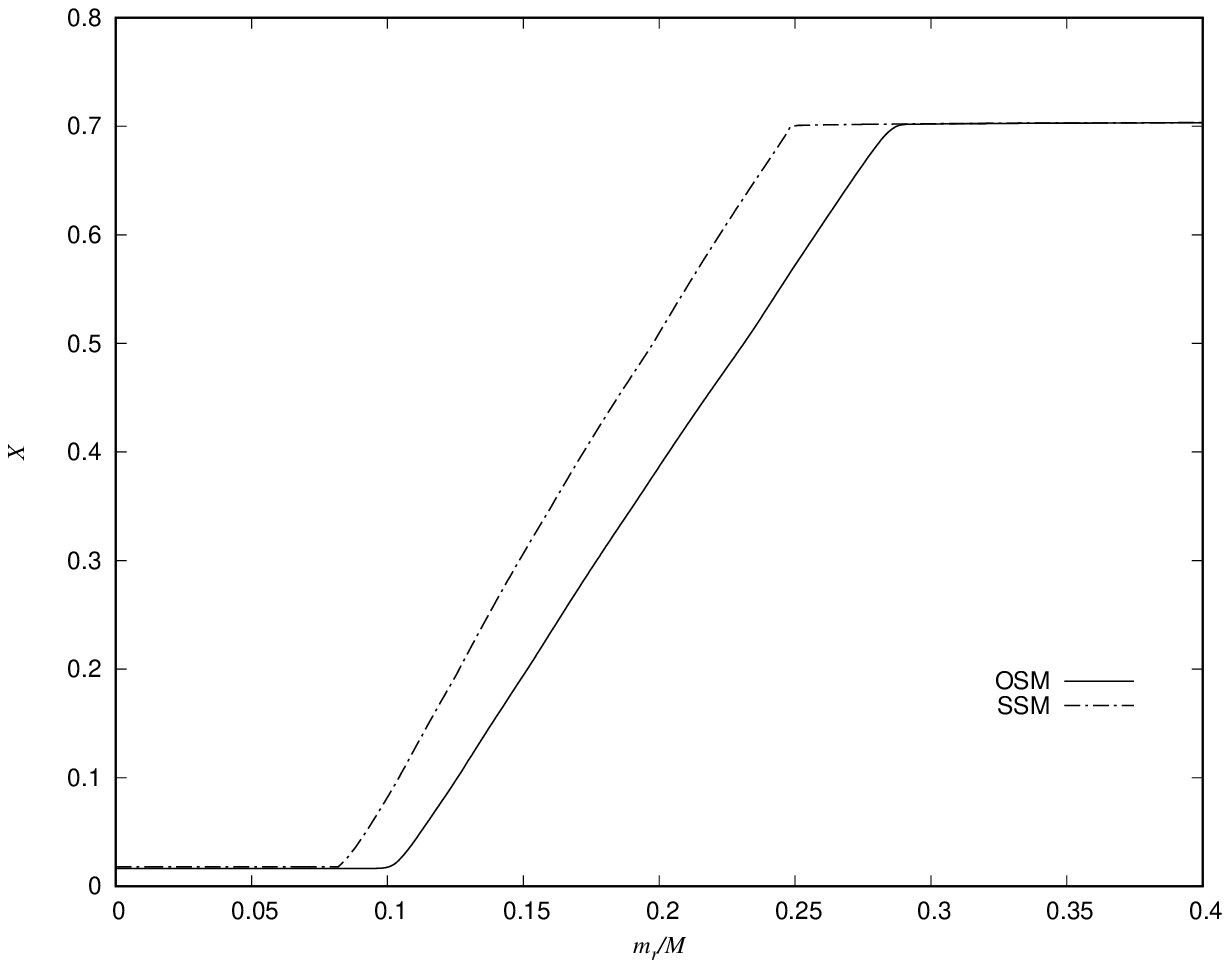}
\caption{Distributions of the hydrogen abundance for the standard and overshooting models of a 5$M_{\odot}$ main sequence star. }
\end{figure}

\begin{figure}
\plotone{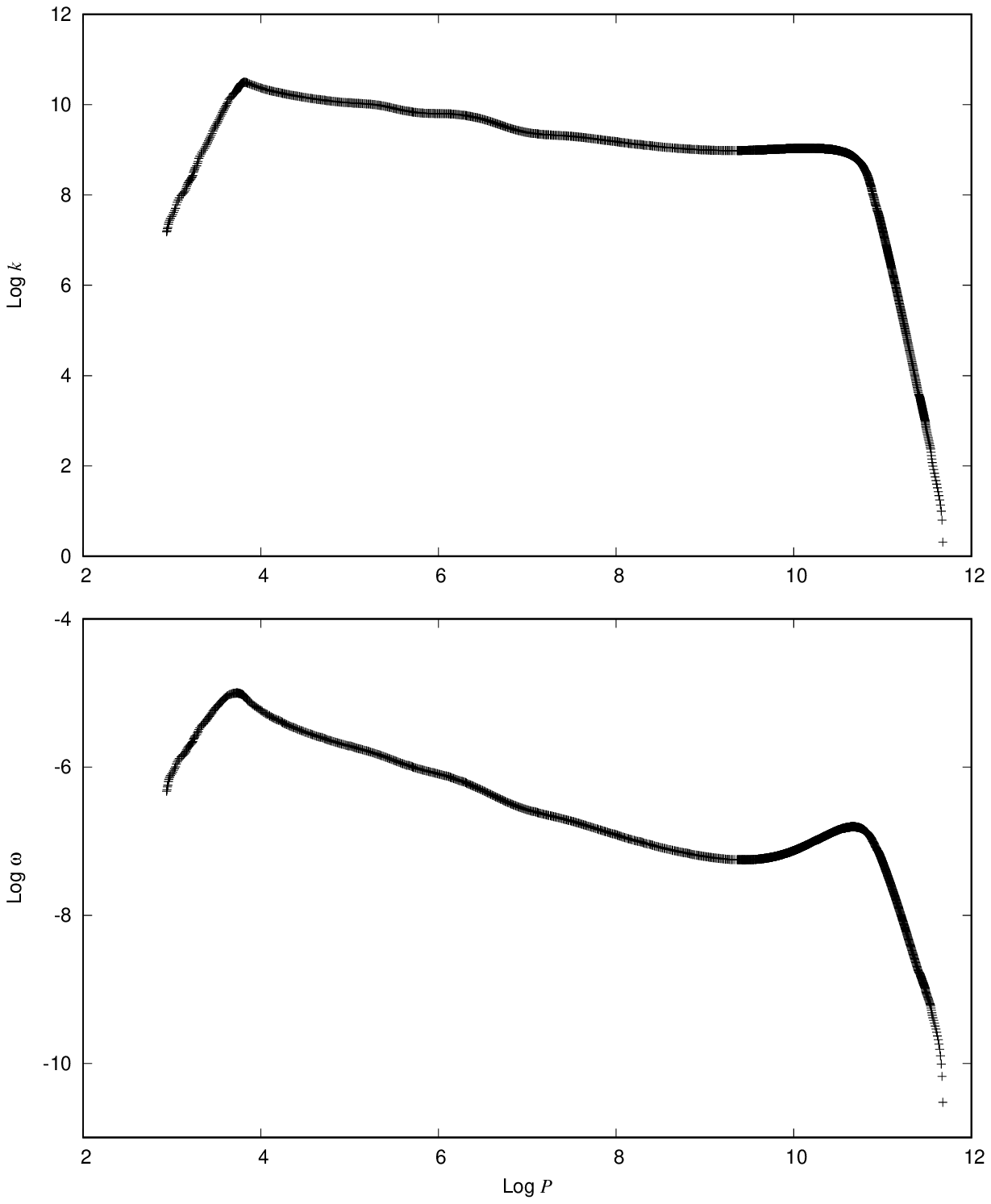}
\caption{Distributions of the turbulent kinetic energy (upper panel) and turbulence frequency (lower panel) for a 5$M_{\odot}$ RGB model. }
\end{figure}

Distributions of the turbulent kinetic energy and the turbulence frequency are shown in Figure 10 for a stellar model on the red giant branch. It can be seen that they have similar behaviors as in the solar case. Figure 11 shows profiles of the hydrogen abundance for some overshooting stellar models during the first dredge-up phase. Instead of discontinuous distributions resulted from the mixing-length type convective mixing treatment in the standard stellar models (SSM), our overshooting stellar models show still steep though smooth hydrogen profiles near the base of the convective envelopes. 

\begin{figure}
\plotone{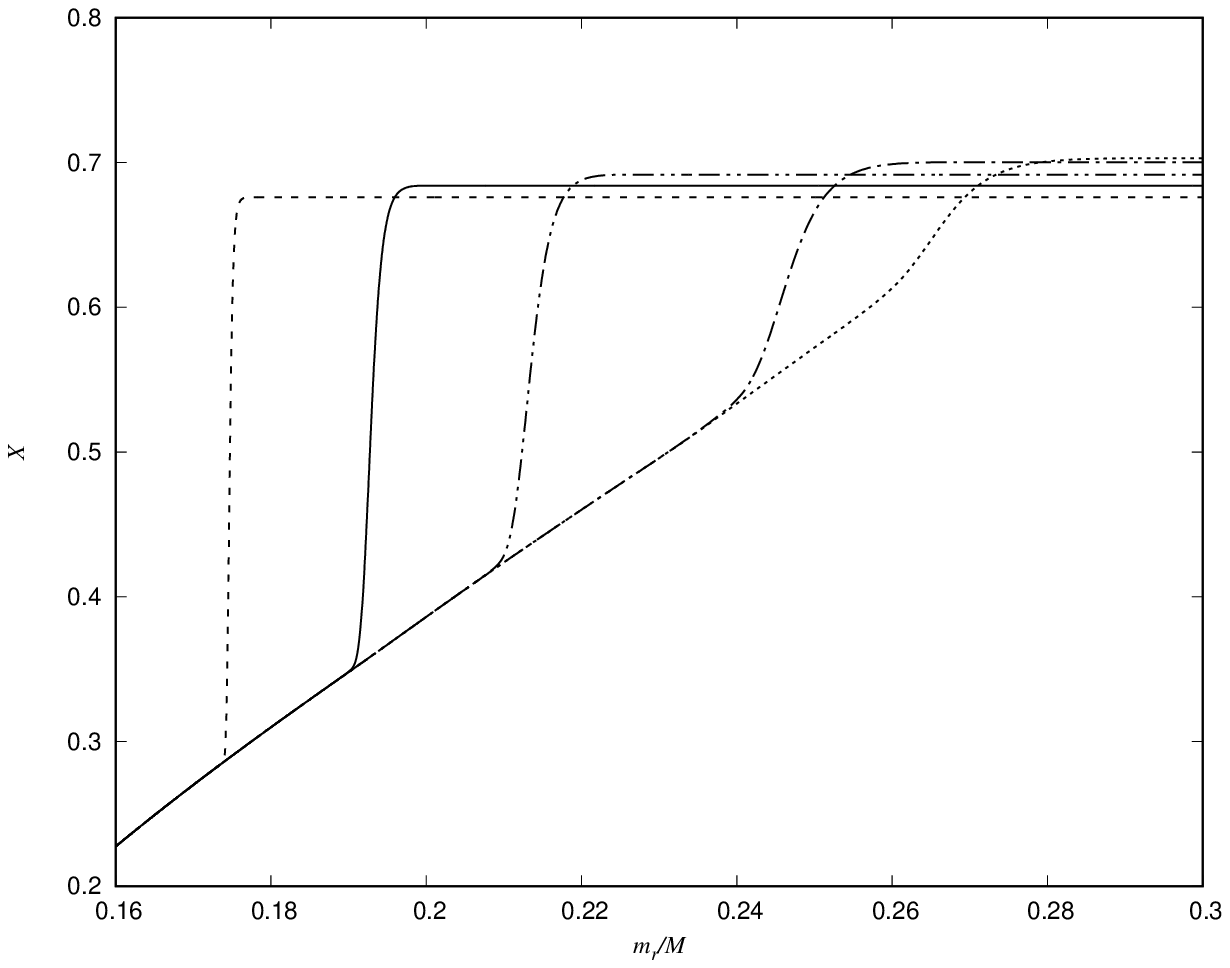}
\caption{Distributions of the hydrogen abundance for the overshooting models of a 5$M_{\odot}$ star during the RGB phase. }
\end{figure}

We compare results of stellar evolution including either the core overshooting during the main sequence or the envelope overshooting during the red giant branch (RGB) in Figure 12. It can be seen that including the envelope overshooting alone does not change the stellar evolution as compared to the standard stellar models, except for a considerably more extended blue loop. This result can be understood by realizing the fact that the development of a blue loop is a direct result of the hydrogen burning shell encountering the chemical discontinuity left by the convective dredge-up before in the RGB stage. The increment of the hydrogen abundance across the chemical discontinuity results in an increment of the opacity, which forms a barrier for the heat transfer. When the hydrogen burning shell get closer to the barrier, more and more heat will be blocked, which will warm the barrier itself to higher and higher temperature. The increment of the temperature leads to the decrement of the opacity and the dispersal of convection in the hydrogen abundant envelope, which are responsible for the blueward evolution of the star. Therefore the overshooting below the base of the convective envelope moves the chemical discontinuity a little bit inwardly, and brings more helium into the stellar convective envelope to make it more transparent for radiation to flow through. These two effects help the star to form a more extended blue loop. 

\begin{figure}
\plotone{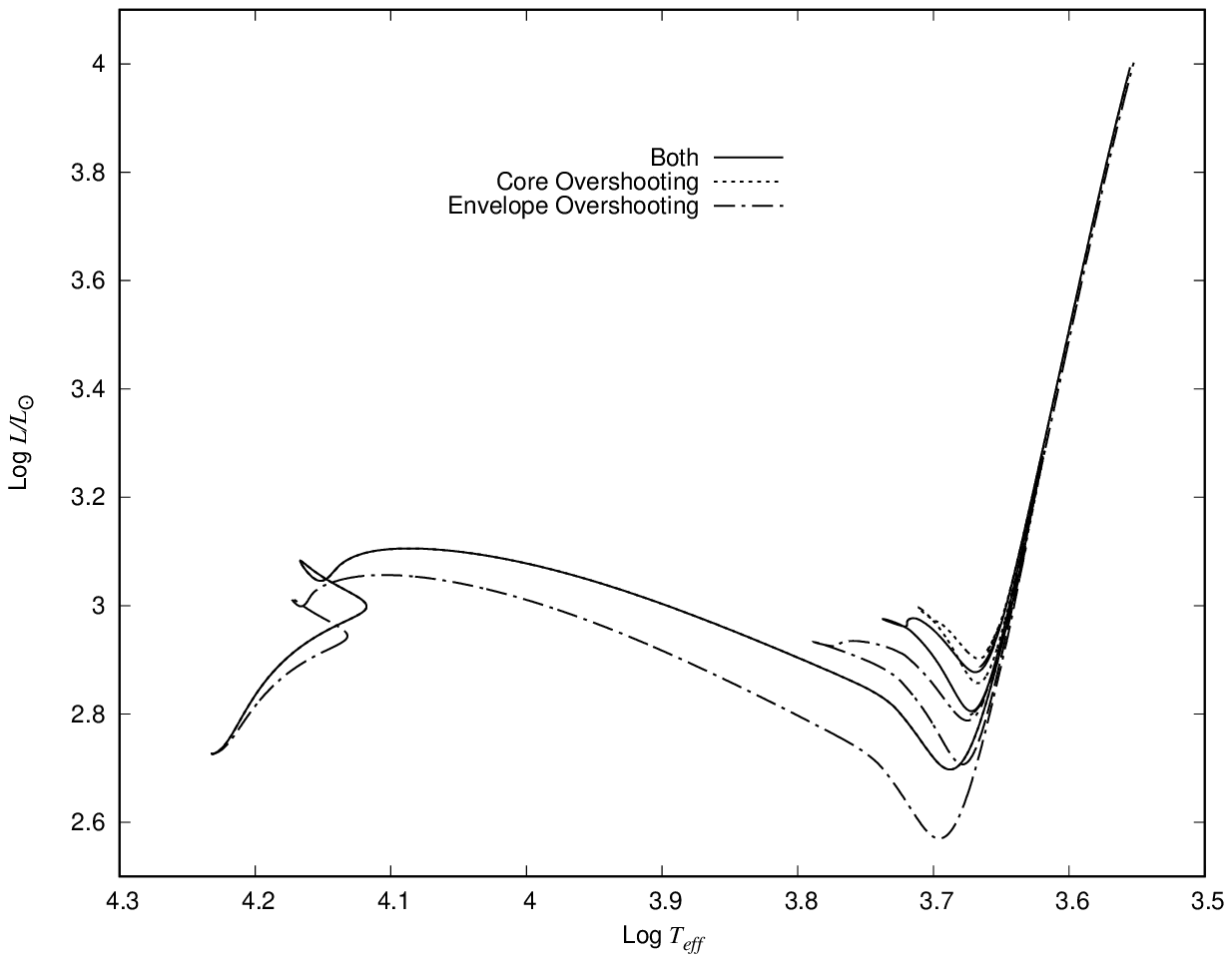}
\caption{Evolutionary tracks of a 5$M_{\odot}$ star taking into account the convective core overshooting in the main sequence and the convective envelope overshooting in the RGB phase. }
\end{figure}

On the other hand, including the core overshooting alone results in higher luminosities for stellar models during central hydrogen and helium burning stages, but leads to a shorter blue loop as shown in Figure 12. These results can be understood as follows. Including the convective core overshooting in the main sequence will enlarge the convectively mixing core and then increase the central temperature. As a result, the star will have a higher luminosity and a longer lifetime in the main sequence as compared with the standard stellar models. When hydrogen is depleted in the central core, the star will develop a helium core and evolve onto the red giant branch. It is important to realize that the convective dredge-up during the RGB stage is actually ended by the ignition of helium at the stellar center. Thus the larger the helium core is, the higher the temperature will be at its center and the earlier the convective dredge-up will be stopped. Therefore including the convective core overshooting in the main sequence will result in a shallower chemical discontinuity, leading to a shorter blue loop afterwards as seen in Figure 12. It can be noticed additionally that including both core and envelope overshooting results in almost the same blue loop as the standard stellar models, but significantly higher luminosities.

What are the effects of parameters of the $k$-$\omega$ model on the evolution of the $5M_{\odot}$ star ? The most important parameter is ${\alpha}'$, which directly determines the size of the convectively burning core during the main sequence. The larger its value is, the bigger the convectively burning core and the higher the stellar luminosity will be. Another important parameter is $\alpha$, which determines the distance of the overshooting mixing below the base of the convective envelope during the helium burning phase. The most significant effect is the extension of the blue loop. Numerical experiments showed that the larger the value of $\alpha$ was, the more extendedly the blue loop would develop.

\section{ Conclusions and Discussions }

The $k$-$\omega$ model for stellar convection was proposed by \citet{Li2012}. We applied it to calculations of various stellar models, and made substantial improvements over the original model itself and the parameters it introduced. In this paper we present the revised $k$-$\omega$ model and show some applications. The improvements on the $k$-$\omega$ model are summarized as follows. 

\begin{enumerate}

\item We have modified the macro-length model of turbulence by discriminating two distinct situations, i.e. full convective envelopes where the thickness of the convection zone is much larger than the local pressure scale height and convective cores where the thickness of the convection zone is usually smaller than the local pressure scale height. 

\item We have modified the convective heat flux according to Kays-Crawford model.

\item We have proposed a model for the convective mixing flux, in accordance with the model for the convective heat flux.

\end{enumerate}

We have applied the improved $k$-$\omega$ model in calculations of solar models. The results show that there are two overshooting regions beyond the solar convective envelope, one below its base boundary and the other above its top boundary. Turbulence decays in both overshooting regions according to the power law. Below the base of the convection zone, there is a mixing-length type overshooting region with the adiabatic temperature gradient and complete mixing, and a diffusion type overshooting region with a sub-adiabatic but super-radiative temperature gradient and partial mixing. On the other side, overshooting develops below the top of the convection zone and penetrates vigorously outward into the solar atmosphere, keeping the chemical composition to be the same as in the convective envelope. 

We have calculated evolutionary models of a 5$M_{\odot}$ star using the revised $k$-$\omega$ model. An overshooting region beyond the convective core is found for stellar models in the main sequence phase. Again, turbulence decays in the power law, resulting in a partial mixing in the overshooting region. Stellar models with the convective overshooting effect are brighter than those without considering the overshooting, and the overshooting models also evolve to lower effective temperatures to define a redder boundary of the main sequence. During the red giant phase, the stellar models develop a convective envelope with an overshooting region below, which is similar to the solar case. The most significant effect of the envelope overshooting is to expand significantly the extension of the blue loop. On the other hand, the core overshooting in the main sequence phase acts to impede the development of the blue loop. These two effects cancel out more or less each other, leaving an almost unchanged blue loop as compared to stellar models without considering the overshooting effects.

We have developed an independent Fortran module to solve the equations of the $k$-$\omega$ model by use of the Newton iterative method. This module is available for any interested researcher when requested.

For hot stars, there are several thin convection zones in their envelopes that can become smaller than the local pressure scale height. Our $k$-$\omega$ model cannot handle this situation at present. It should also be noted that the turbulent pressure is neglected in the equation of hydrostatic support for all of our stellar models. These drawbacks should be solved step by step in the future work.

\acknowledgments 

We cordially thank an anonymous referee for his productive comments, which
greatly help us to improve the manuscript. This work is funded by the NSFC of China (Grant No. 11333006 and 11521303) and the foundation of Chinese Academy of Sciences (Grant No. XDB09010202).

{}

\clearpage

\clearpage

\end{document}